\documentclass{ifacconf}
\usepackage{graphicx}      % include this line if your document contains figures
\usepackage{natbib}        % required for bibliography
\usepackage{amsmath}
\usepackage{amsfonts}
\usepackage{subcaption}
\usepackage{graphicx}
\usepackage{multirow}
\usepackage{amssymb}
\usepackage{makecell}
\usepackage{url}
\newcommand{\B}[1]{\mathbf{#1}}

\begin{document}
\begin{frontmatter}

\title{Baseline Results for Selected Nonlinear System Identification Benchmarks} 
% Title, preferably not more than 10 words.

\thanks[footnoteinfo]{M. Champneys and T. Rogers were supported by the EPSRC grant EP/W002140/1. \\ M. Schoukens: This work is funded by the European Union (ERC, COMPLETE, 101075836). Views and opinions expressed are however those of the author(s) only and do not necessarily reflect those of the European Union or the European Research Council Executive Agency. Neither the European Union nor the granting authority can be held responsible for them.}

\author[First]{Max D. Champneys}  
\author[Second]{Gerben I. Beintema}  
\author[Second,Third]{Roland. Tóth} 
\author[Second]{Maarten Schoukens}  
\author[First]{Timothy J. Rogers}  

\address[First]{Dynamics Research Group, University of Sheffield, Sheffield, UK}
\address[Second]{Control System group, Eindhoven University of Technology, Eindhoven, the Netherlands}
\address[Third]{Systems and Control Laboratory, Institute for Computer Science and Control, Budapest, Hungary}

% why benchmark?
% - Facilitate access to datasets 
% - test approaches against a range of nonlinear phenomena
% - permit an objective comparison of different methods

% why baseline?
% - Establish minimum performance levels against which advanced methodologies might be compared.
% - Prevent 'isolated comparisons' within model classes that do not compare to basic linear identification techniques.
% - 

\begin{abstract} % Abstract of not more than 250 words.
Nonlinear system identification remains an important open challenge across research and academia. %
Large numbers of novel approaches are seen published each year, each presenting improvements or extensions to existing methods. %
It is natural, therefore, to consider how one might choose between these competing models. %
Benchmark datasets provide one clear way to approach this question. %
However, to make meaningful inference based on benchmark performance it is important to understand how well a new method performs comparatively to results available with well-established methods. %
This paper presents a set of ten baseline techniques and their relative performances on five popular benchmarks. %
The aim of this contribution is to stimulate thought and discussion regarding objective comparison of identification methodologies.
\end{abstract}


\end{frontmatter}

\section{Introduction}

Nonlinear system identification (NLSI) remains an open challenge in all but a few restricted cases. %
A testament to the difficulty of identifying nonlinear systems is the wide variety of techniques still being published in the literature, so much so, that it sustains several dedicated conferences and workshops. %including this IFAC Symposium on the topic of system identification, where many contributions focus on nonlinear systems. %
Furthermore, and perhaps unsurprisingly, as in many areas, the system identification community is embracing machine learning (ML) techniques to address open problems in the field. %
The goals of machine learning; to automatically uncover unknown mappings from data, align very closely with the desires of the system identification practitioner. %

In the midst of numerous methodological contributions, the question remains: ``Which is the best approach?'', or more pertinently: ``Which approach should be used in my application?''. %
Unfortunately (or fortunately for those whose career is in this research field), the answer is highly dependent on the individual problem at hand. %
Going some way to addressing this ambiguity, as is seen in ML settings, several benchmark datasets have emerged to allow for meaningful comparison between approaches, see \citep{Schoukens2009,Wigren2013,SchoukensM2016,SchoukensM2017, Wigren2017, Janot2019} among others\footnote{\url{nonlinearbenchmark.org}}. %
Benchmark datasets provide a point of objective comparison where methods cannot only be demonstrated in restricted scenarios, chosen by the authors to promote their novel approaches. %
Access to high-quality experimental datasets also enables researchers to develop and test methods against more data than might otherwise be available; reducing the burden of data collection alongside development of novel identification methods. %

\subsection{The Need for Baselines in NLSI}
Where benchmarks enable a direct comparison between modelling approaches, evaluation of performance metrics additionally offer insight into the relative performance of models. %
For example, the mean squared error (MSE) score is a convenient metric by which both similar and dissimilar methods can be compared objectively. %
In this paper, the authors make the claim that alongside benchmark datasets, it is imperative that \emph{baseline} results are reported. %
This need is motivated by the rapid emergence of ML technologies being developed and applied in NLSI.  %
Baseline results for well-established approaches defend against isolated comparisons between models from a single class. % 
For example, when proposing novel neural-network architectures for NLSI, comparisons may be made to existing architectures (e.g. a simple feed-forward network) but neglect to show comparisons to a linear dynamic model. % 
By no means is this issue universal; there are many good examples in the literature where thorough comparisons are made. In cases where baseline comparisons are absent, the authors do not believe that this is a malicious omission. %
Instead, it speaks to the difficulty of obtaining (by implementation or review) baseline results for many NLSI models alongside those being presented.  % 

The argumentation of the authors is that the benefit of benchmarks datasets is lost when comparisons made on these benchmarks do not acknowledge that breadth of modelling approaches available in the literature. %
The role of this paper is therefore to highlight the quality of results which may be obtained from existing classical models, such that, future contributions can be reasonably compared. % 
This work will also highlight, from among the benchmarks considered, which datasets pose particular challenges for certain model structures. %

In the next sections, we will briefly introduce the considered benchmark systems (Section~\ref{sec:Benchmarks}), followed by an outline of the considered identification approaches in Section~\ref{sec:Approaches}. Next, Section~\ref{sec:Results} shows the obtained baseline results and they are discussed in more depth in Section~\ref{sec:Discussion}.

\section{Benchmark Systems} \label{sec:Benchmarks}%@MS 
% Maybe we can do this as a paragraph rather than as subsections? -MDC  

For the purposes of comparison between nonlinear system identification methods, a number of benchmark systems have been proposed\footnote{A large number of these are collected at \url{nonlinearbenchmark.org}}. %
From this set of benchmark systems several are more regularly used for demonstration than others, these are typically the SISO systems, and the datasets that have a more rich input. In this paper, five of these benchmarks are used to demonstrate possible results with baseline models. For each of these, a  Python data-loader is available online\footnote{\url{github.com/GerbenBeintema/nonlinear_benchmarks}}, that automates the division of the benchmark data into the same training (incl. validation), and testing sets that are used in this study. The benchmarks used in this paper are: the Silverbox \citep{Wigren2013} which mimics a Duffing oscillator/Fitz-Nagumo model, the EMPS system \citep{Janot2019} which exhibits friction behaviour, the Wiener-Hammerstein (W-H) system \citep{Schoukens2009} containing a saturation nonlinearity between two LTI blocks, the cascaded tanks (CT) system \citep{SchoukensM2016,SchoukensM2017} with overflow nonlinearities, and the coupled electric drives (CED) \citep{Wigren2017} wherein only the absolute value of the velocity is measured. Full details can be found in their respective references.
 
\section{Identification Approaches} \label{sec:Approaches}

% - Showcase well-established methods 
% - Attempt to encompass mayor axes of NLSI models (SS, NARX, NN)
% - Not over-specialise approaches to the datasets
% - MPO/Simulation only

It is the intention of the authors to demonstrate baseline results on well-established and readily available methodologies. %
The modelling approach here is to apply models without significant modification or tailoring to each benchmark problem. %
This is motivated by a desire to showcase results that might be expected given a non-intrusive application of commodity algorithms. %
As is well known in the NLSI community, the most rigorous test of model performance is in free-running simulation (where only the initial conditions of the system output are observed) \citep{Schoukens2019}. %
Therefore, all models are tested exclusively in simulation. % 
In all cases where data normalisation is applied, this is calculated using the training data only. All error metrics are reported in the original coordinates of the benchmarks. %
In this paper, baseline NLSI models are compared for three common model classes; state-space models (SSM), auto-regressive models (AR) and recurrent neural networks (RNN). %
Each of these is discussed briefly below. %

\subsection{LTI State-Space}
Discrete-time LTI state-space models have been commonly used for decades in the system identification literature. Here, LTI state-space models with an output error noise model are estimated:
\begin{align} \begin{aligned}
    x_{t+1} &= A x_t + B u_t \\
    \hat{y}_{t} &= C x_t + D u_t
\end{aligned}\end{align}
where $u_t$ is the input, $x_t$ is the state, $\hat{y}_t$ is the modeled output at time $t$. The matrices $A$, $B$, $C$, and $D$ are parameterised in a `free' form, i.e. any element in the matrix is freely adjustable. The state-space models are estimated using the prediction error framework. The resulting nonlinear optimization problem is initialised using a subspace estimate \citep{Ljung1999}.

The means of the measured input and output training data are subtracted before estimating the LTI state-space models. Hence, minimizing the prediction error results in a Best Linear Approximation model of the considered systems, following the lines of \citep{Enqvist2005a,Pintelon2012}. The resulting mean values of the input and output training data are part of the resulting model as they are respectively subtracted from and added to any validation or test input provided to the model and resulting model output.

The state order $n_x$ is selected based on the provided benchmark system descriptions, except for the EMPS benchmark, were a grid search with validation dataset illustrated that $n_x = 3$ resulted in significantly better results than $n_x=2$.

\begin{table}
    \centering
    \caption{Number of states used by the SS models considered in this paper.}
    \begin{tabular}{c||ccccc}
         &  &  &  & Cascaded &  \\
        Benchmark & Silverbox & W-H & EMPS & Tanks & CED \\ \hline
        $n_x$ & 2 & 6 & 4 & 2 & 3
    \end{tabular}
    % \begin{tabular}{c||cc}
    %     Benchmark &  $n_x$  \\ \hline
    %     Silverbox & 2  \\
    %     W-H       & 6 \\
    %     EMPS      & 3  \\
    %     Cascaded Tanks & 2  \\
    %     CED       & 3  \\
    % \end{tabular}
    \label{tab:SSO}
\end{table}

\subsection{Auto-regressive models}

Auto-regressive (AR) models \citep{Billings2013,Ljung1999} encompass a class of methods that generate predictions of output variables given previous (or \emph{lagged}) values. Many variants of AR models are well established in the NLSI literature, generalising the linear, output-only AR formulation to exogenous inputs (ARX), nonlinear systems (NARX) and moving-average noise processes (ARMA). %
A convenient general representation of nonlinear auto-regressive models is to consider the explicit prediction of the the subsequent time point $\hat{y}_{t}$ as, 

\begin{equation}
    \hat{y}_{t} = \Phi(y_{t-1}, \ldots, y_{t-n_y}, u_t, \ldots, u_{t-n_u+1})
\end{equation}

where $\Phi$ is a static map from the lagged inputs and outputs to the predictions, and where $n_u$ and $n_y$ refer to the number of lagged inputs and outputs respectively. This has the advantage of transforming the system identification procedure into a static regression problem, to which many standard approaches can be applied. A convenient method of working with AR type models is to construct the Hankel matrix of lagged inputs and outputs,

\begin{equation}
   H= \begin{bmatrix}
        y_{t-1}   & \ldots & y_{t-n_y}   & u_t          & \ldots & u_{t-n_u +1} \\
        y_{t-2}   & \ldots & y_{t-n_y-1} & u_{t-1}      & \ldots & u_{t-n_u }   \\
        y_{t-3}   & \ldots & y_{t-n_y-2} & u_{t-2}      & \ldots & u_{t-n_u -1} \\
        \vdots    &        & \vdots      & \vdots       &        & \vdots       \\
        y_{t-N+p} &        & y_N         & u_{t-N+p-1}  &        & u_{N-1}      \\
    \end{bmatrix}
\end{equation}

where $N$ is the total number of temporal points in the data and $p=\operatorname{max}(n_y, n_u)$. This formulation allows one to express the prediction model as $\B{y} = \Phi(H)$, where the static function $\Phi$ depends on the specific form of the auto-regressive model under consideration. An open problem in AR modelling is how to specify the number of lags $n_y$ and $n_u$. %
A cross-validation approach is often cost-prohibitive due to the combinatorial nature of the lag structure problem. %
The approach taken in this work, for all AR models detailed below, is to choose the lag structure with the best score for a linear ARX model, evaluated by the Akaike information criterion (AIC) \citep{akaike1998information} on a hold-out validation (not testing) set, up to a maximum value of $n_y, n_u = 20$. %
The only exception is for the CED benchmark for which this approach selects no lags on the output ($n_y =0 $), likely due to the symmetric nonlinearity on the output resulting in a minima at zero (in the least-squares sense) for input signals with zero mean \cite{SchoukensM2017}. To avoid disadvantaging the nonlinear models, the lag structure for this benchmark is set (arbitrarily) to $n_y, n_u = 10$. %
The lag structures of the AR models reported in this study are collected for the convenience of the reader in Table \ref{tab:LSO}. %

\begin{table}
    \centering
    \caption{Lag structures used by the AR models considered in this paper.}
    \begin{tabular}{c||ccccc}
         &  &  &  & Cascaded &  \\
        Benchmark & Silverbox & W-H & EMPS & Tanks & CED \\ \hline
        $n_x$ & 10 & 8 & 16 & 9 & 10\\
        $n_y$ & 10 & 15 & 5 & 8 & 10\\
    \end{tabular}
    % \begin{tabular}{c||cc}
    %     Benchmark &  $n_u$ & $n_y$ \\ \hline
    %     Silverbox & 10 & 10 \\
    %     W-H       & 8 & 15\\
    %     EMPS      & 16 & 5 \\
    %     Cascaded Tanks & 9 & 8 \\
    %     CED       & 10 & 10 \\
    % \end{tabular}
    \label{tab:LSO}
\end{table}

\subsubsection{ARX:}%
The first auto-regressive model considered here is the auto-regressive with exogenous inputs (ARX) model. This is a linear formulation with,

\begin{equation}
    \Phi(H) = H \alpha
    \label{eqn:ARX}
\end{equation}

where $\alpha$ is an unknown vector of linear weights that can be evaluated by the solution of the linear equation $\B{y}= H\alpha$. Many solvers are appropriate for this task. The results in this paper use a QR-decomposition based approach. 

\subsubsection*{pNARX:}
The first nonlinear ARX (NARX) model considered in this study is the classical polynomial NARX model \citep{Billings2013}. Here, the nonlinearity is defined as linear combination of polynomial terms $P_n$ computed from $H$. %

\begin{equation}
    \Phi(H) = P_n(H) \alpha
\end{equation}

In order to promote the stability of the modelling approach, only a monomial basis is considered (cross terms such as $y_{t-1}^2 u_{t-1}$ etc. are ignored). %
Furthermore a Legendre orthogonal polynomial basis is used to improve the numerical conditioning. %
The order of the polynomial is selected based on AIC cross-validation scores on the hold-out validation set, up to a maximum order of 7. %
It is expected that the inclusion of multinomial terms may improve the identification on some benchmarks. %
However, a full multinomial basis leads to a very high number of basis terms for only a modest number of lags. %

\subsubsection*{GP NARX:}

Another common choice for the nonlinearity in NARX models is to use a Gaussian process (GP) model. GPs offer flexible kernel regression and quantification of uncertainty with few hyperparameters \citep{Rasmussen2006}. %
The nonlinearity is therefore,

\begin{equation}
    \Phi(H) \sim \mathcal{GP}\left(m(K), K(H, H)\right)
\end{equation}

The GP is entirely parameterised by the kernel function $K$ and mean function $m$. In this study, GP NARX models are constructed with a squared-exponential kernel and zero-mean function. %
Training GPs requires the specification of several kernel hyperparameters $\theta_K = \{\sigma_n^2 \sigma_f^2, \ell\}$. %
By convention, these are optimised by minimisation of the negative-log marginal-likelihood with respect to the training data \citep{Rasmussen2006}. %
For a zero-mean GP, this is given by,

\begin{equation}
   - \log p(\theta_K | \B{y}) \propto  \frac12 \B{y}^T\Sigma^{-1}\B{y} + \frac12 |\Sigma|  % TODO check @MDC
\end{equation}

with $\Sigma = K(H, H) + \sigma^2_n I$, for the proposed noise variance term $\sigma_n^2$.

Here, this is achieved with the well-known stochastic gradient descent algorithm ADAM \citep{kingma2014adam}. %
For each benchmark, the hyperparameters are optimised from random initialisations for a total of 1000 steps with a learning rate of $10^{-2}$. %
The hyperparameters with the best score on the hold-out validation set are then used for evaluation of the test set. %

Due to the computational cost ($\mathcal{O}(n^3)$ for $n$ datapoints), of training GPs, for benchmark sets with more than 1000 examples (EMPS, W-H, SB), only the first 1000 points of the training set are used to train the model.  %

\subsubsection*{MLP NARX:}

The final nonlinearity considered for NARX models in this study is to use a multi-layer perceptron (MLP) for $\Phi$. %
Here, a MLP with a single hidden layer and hyperbolic tangent activation functions is considered. %
The number of nodes in the hidden layer is selected by cross validation in the hold-out validation set with sizes of 2, 5, 7 and 10 considered.  % 
Each network is trained for $2\times10^4$ iterations of the ADAM \citep{kingma2014adam} optimiser, with 5 random initialisations. %
The overall best network is selected as the one with the best score on the hold-out validation set.  %
This network is then used for evaluation on the unseen testing set. % 

\subsection{Recurrent neural-networks}

In recent years, the field of NLSI has benefited greatly from significant developments in the realm of ML. In particular, recurrent neural network (NN) models have gained a lot of attention within the system identification community and variants have generated impressive results on challenging benchmarks.
RNNs approximate the evolution of dynamic systems by propagating a hidden `memory' state (or states) forward through the time evolution of the network. %

This extends the standard MLP formulation to enable dynamical systems to be modelled in an input-output fashion.% 
It is common when training RNNs to stack lagged values of the inputs up to a certain \emph{look-back} distance.%
This can be achieved in a similar way to the AR models by constructing the Hankel matrix of the input values only. %
The network is then trained to predict the value of the output and the next sampling instant from the lagged inputs and the previous value of the hidden memory state. %  

\begin{equation}
    \begin{bmatrix}
    \B{h}_{t+1} \\
    \B{y}_{t+1}
    \end{bmatrix} =  \B{F} (\B{h}_{t}, u_{t+1}, \dots u_{t-n_u+1})
\end{equation}

Where $\B{F}$ is the forward pass of the neural network and $n_u$ is the look-back distance. In the RNN literature $\B{F}$ is often called a \emph{recurrent cell}. %
Complex RNN models are often comprised of many such cells composed in series and parallel. %

Although many advanced architectures have been proposed in a NLSI context \citep{beintema2021nonlinear, beintema2023continuoustime, forgione2021dynonet, forgione2021continuous}, the objective of this paper is to compare baseline results for standard architectures; as such, this study considers only models with a single recurrent cell. %

In all cases, the look-back distance and the dimension of the hidden memory state are selected by cross-validation on the hold-out validation set, with the best parameterisation selected for evaluation on the testing set, these values are not exposed here for reasons of space but are available in the data presented alongside this paper\footnotemark[4]. %
Training is conducted in batches, using 10 random initialisations of the ADAM (with learning rate parameter $10^{-3}$) optimiser for between $2\times10^4$ and $10^4$ epochs depending on the amount of data available for training. %
For reasons of space, an explicit description of the various RNN architectures is not presented here. %
The interested reader can see \citep{pml1Book} (or many other good textbooks) for additional details on the models used in this study. %

\subsubsection*{MLP FIR:}

The first structure considered in this study is a (nonlinear) finite impulse response (FIR) model. This model structure is equivalent to removing the hidden `memory' state from the recurrence relation above. 
\begin{equation}
   \B{y}_{t+1} =  F(u_{t+1}, \dots u_{t-n_u+1})
\end{equation}

Where $F$ is an MLP with a single hidden layer. Although this architecture is not recurrent, it will serve to highlight the extent to which a recurrent formulation improves the identification for each of the benchmark systems.

\subsubsection*{RNN:}

The canonical formulation of an RNN \citep{schuster1997bidirectional} constructs the hidden cell as a single dense layer in an MLP, with a hyperbolic tangent activation function --- other choices are clearly available. %
This architecture is employed here with the size of the hidden layer decided by cross-validation as described above. %

\subsubsection*{GRU:}

A known issue with the RNN is that as the gradient is propagated backwards in time, successive applications of the chain rule over activation functions can cause the numerical values of the gradients to become arbitrarily small or large. %
This issue is known as \emph{vanishing} (resp. \emph{exploding}) gradients \citep{pascanu2013difficulty}. Several variants of the basic RNN cell have been proposed to try and combat this issue, by applying a `gating' function (often a sigmoid or similar) to the value of the memory state in each iteration. %
The gated recurrent network (GRU) \citep{chung2014empirical} is one such architecture that employs this technique.  %

\subsubsection*{LSTM:}

The original extension to the RNN is the famed \emph{long-short term memory} (LSTM) unit \citep{hochreiter1997long}. This architecture propagates an extended hidden state consisting of both `long-term' and `short-term' memory. %
The \emph{optimised LSTM} (OLSTM) is a variant of the LSTM cell that concatenates the hidden states prior to the activation functions for efficient computation. %
Results from both the LSTM and OLSTM networks are reported in this study. %

\section{Baseline Results} \label{sec:Results}

In order to establish the baseline results all of the models described above are evaluated on the unseen testing set. %
For comparison, all model performances are reported using the root mean-squared error metric,
\begin{equation}
    J_{\text{RMS}} = \sqrt{\frac{1}{N} ||y- \hat{y}||^2}    
\end{equation}

All model performances are reported in Table \ref{tab:BaselineResults}\footnote{Code to reproduce these baseline results can be found at \url{github.com/MDCHAMP/nonlinear_baselines}}. Note that in order to maintain readability, the reported error scores are not all presented in the same units as the benchmarks. In particular, the SB, W-H, and EMPS benchmarks have been scaled by a factor of $10^{-3}$. To put these numbers in perspective, a selection of state-of-the-art (SOTA) results are collected in Table \ref{tab:SOTAResults}.
\begin{table*}[ht!]
    \centering
    \caption{Baseline simulation RMSE results on benchmark datasets.}
    \label{tab:BaselineResults}
    \begin{tabular}{c||ccc|c|c|c|cc} 
         Benchmark&  \multicolumn{3}{c|}{Silverbox}  &  W-H &    EMPS&   \makecell{Cascaded \\ Tanks} &  \multicolumn{2}{c}{CED}\\ 
         \makecell{Figure of Merit \\ (Unit)} &  \multicolumn{3}{c|}{RMS (mV)}  &  RMS (mV) &  RMS (mm)&  RMS (V)&  \multicolumn{2}{c}{RMS (ticks/s)} \\ \hline 
         Test Set& multisine & \makecell{ arrow \\ (full)} & \makecell{ arrow \\ (no extrap.) }  & test & test & test & test 1 & test 2 \\ \hline 
         LTI SS    & 6.96   & 14.4  & 6.58  & 43.4 & 5.53  & 0.588  & 0.224 & 0.384 \\ \hline
         % PNLSS     &  &  &  &  &  &  &  &  \\ \hline  

         LTI ARX	  & 6.95   & 14.2  & 6.59  & 43.4 & 12.6  & 0.685  & 0.291 & 0.331 \\
         pNARX	  & 0.640  & 2.25  & 0.571 & 27.3 & 5.17  & 0.413  & 0.10  & 0.141 \\
         GP NARX	  & 0.301  & 0.60  & 0.259 & 23.1 & 1690  & 0.622  & 0.102 & 0.090 \\
         MLP NARX  & 2.80   & 5.18  & 2.40  & 310  & 133   & 2.04   & 0.142 & 0.101 \\ \hline

         MLP FIR   & 18.9   & 20.2  & 15.3  & 24.8 & 121   & 1.43   & 0.122 & 0.095 \\ 
         RNN       & 1.64   & 4.88  & 1.73  & 6.87 & 89.2  & 0.543  & 0.096 & 0.235 \\  
         GRU       & 1.49   & 2.80  & 0.947 & 3.97 & 220   & 0.396  & 0.286 & 0.238 \\  
         LSTM      & 1.50   & 2.68  & 0.960 & 4.45 & 74.1  & 0.490  & 0.303 & 0.145 \\
         OLSTM     & 1.51   & 2.38  & 1.08  & 4.55 & 111   & 0.471  & 0.151 & 0.259 \\ \hline

    \end{tabular}
\end{table*}

\begin{table*}[ht!]
    \centering
    \caption{Selected state-of-the-art results available in the literature.}
    \label{tab:SOTAResults}
    \begin{tabular}{c||ccc|c|c|c|cc}
        Reference & \multicolumn{3}{c|}{Silverbox} & W-H              & EMPS & \makecell{Cascaded                                \\ Tanks} &  \multicolumn{2}{c}{CED}\\
        \makecell{Figure of Merit \\ (Unit)} &  \multicolumn{3}{c|}{RMS (mV)}  &  RMS (mV) &  RMS (mm)&  RMS (V)&  \multicolumn{2}{c}{RMS (ticks/s)} \\ \hline 
        Test Set  & multisine                      & \makecell{ arrow                                                            \\ (full)} & \makecell{ arrow \\ (no extrap.) }  & test & test & test & test 1 & test 2 \\ \hline
        DT subnet \citep{beintema2021nonlinear} & 0.36                           & 1.4              & 0.32 & 0.241              & -    & 0.37  & 0.169 & 0.117 \\
        CT subnet \citep{beintema2023continuoustime} & -                              & -                & -    & -                  & 4.61 & 0.306 & 0.115 & 0.074 \\
        Grey-box \citep{worden2018evolutionary} & -                              & -                & -    & -                  & -    & 0.191 & -     & -     \\
        \makecell{PNLSS \citep{paduart2010identification} \\   \citep{paduart2012identification,relan2017unstructured}} & -                              & 0.26             & -    & 0.42                 & -    & 0.45     & -     & -     \\
        TSEM      \citep{forgione2021continuous} & -                              & -                & -    & -                  & -    & 0.33  & -     & -     \\
        SCI       \citep{forgione2021continuous} & -                              & -                & -    & -                  & -    & 0.4   & -     & -     \\
        dynoNet   \citep{forgione2021dynonet} & -                              & -                & -    & 0.452              & 2.64 & -     & -     & -     \\\hline
    \end{tabular}
\end{table*}

\section{Discussion} \label{sec:Discussion}

Overall, there is significant variation in the performances of the models on the five benchmarks considered in this study. %
It is unsurprising that in almost every case the nonlinear models are outperforming their linear counterparts. %
However, it is interesting to note that for the SB and W-H benchmarks, linear models are nevertheless able to achieve qualitatively very good fits (even if order of magnitude improvements are available). %
In almost every benchmark, the linear models (LTI SS, LTI ARX) achieve near identical scores, though the LTI ARX model typically requires a higher model order to correctly capture the disturbances present in the data. %
This is evidence that the linear models are finding a common linearisation of the nonlinear dynamics. %
The only exception to this trend is that the LTI SS model significantly outperforms the LTI ARX model on the EMPS benchmark.
A possible explanation for the observed performance difference between the linear models is the different model orders (selected by cross-validation) for final evaluation on test set and the simulation focus of the LTI SS model training. %
The presence of a friction nonlinearity in the EMPS benchmark is particularly challenging for linear models to approximate, thus it is possible that the different model orders lead to vastly different linearisations. %
Further evidence of this is seen by setting $n_x = 2$ (as per the physical system) in the LTI SS model. %
In this case, an RMS score of 67.95V is observed --- significantly worse than the LTI ARX result. %

% It is notable that several of the baseline models approach state-of-the-art results on the SB, CED and CT benchmarks. %
% In particular, the pNARX, GP NARX, GRU and LSTM models all offer very competitive results. %
% This serves as an indication of the relative ease of modelling these systems in the black-box setting, however, results on the EMPS and W-H benchmarks lag significantly behind SOTA scores. %

It is notable that most baseline models do not achieve state-of-the-art performance. In some cases though, state-of-the-art results are obtained. In particular, the performance of the pNARX (EMPS, CT, CED), GP NARX (Silverbox, CED) are noteworthy.

The poor performance of most baseline models on the EMPS benchmark (compared to LTI results) indicates that this system offers a robust modeling challenge. %
The presence of friction, and the vastly different lengthscales of the input and output signals contribute to this challenge. %
Several of the models used for evaluation (GP NARX, MLP NARX, RNNs etc.) return predictions with very high error. % 

% Despite a qualitatively good fit to the data, the baseline results on the W-H benchmark remain an order of magnitude away from the SOTA results. %
% A possible explanation for this is the saturation nonlinearity inside the Wiener-Hammerstein block structure. %
% This type of nonlinearity is particularly difficult for black-box learners, despite the linear-dominated dynamics of the W-H system. %

One interesting outcome from the baseline results is the performance of the pNARX model. %
In all benchmarks, the polynomial model demonstrates robust performances (close to or up to a factor 2 worse than SOTA for SB, EMPS, CT and CED). %
Especially remarkable is that these results can be achieved for very little computational effort, even in the presence of large training sequences, and considering only monomial terms in the model \citep{Billings2013}. %
This motivates the use of pNARX as a `first analysis' in the black-box NLSI setting. %
The GRU and LSTM architectures also obtain robust performances on most benchmark systems, though they do not reach SOTA performance. This should maybe not come as a surprise due to the generic nature of these methods, despite well-documented successes in other fields. %
Models that leverage a nonlinear state space structure (PNLSS and subnet models) are also seen to be very performant; giving SOTA results in most benchmarks. %

% While not included as a baseline result, as the model order selection has been executed with more care compared to the baseline models, the nonlinear state space results (subnet, PNLSS), are obtaining the best, or among the best results on all benchmark systems. This emphasizes their flexible nature and helps to motivate their common use in (nonlinear) modelling and identification applications.

The baseline and SOTA results collected in this study also serve to highlight a long-known truism in NLSI: %
The most successful models are those that can leverage both physical insight and data to capture the dynamics. %
A key example of this is the grey-box model of the CT benchmark in \citep{worden2018evolutionary}. %
By leveraging both a physical model of the tanks and the flexible GP NARX model, the authors drive the test RMS score to less than half of that of the most successful baseline, or the physical model on its own. %

%The Silverbox results also offer such insight, as the best results \citep{paduart2010identification} are obtained using a polynomial nonlinearity, as is expected by the underlying system design.

In this paper, extensive model comparisons have been made using the RMS error as a measure of model quality. %
However, it should be noted that the RMS score (or any $\mathcal{L}_p$ distance or any other metric) cannot tell the whole story. %
For example, no consideration has been made here of important factors such as; model complexity, model interpretability, computational complexity, parsimony, stability, or the quantification of uncertainty. %  
Clearly, any successful NLSI approach must consider these factors holistically, and on a case-by-case basis. %
The subject of how to best assess the efficacy of an NLSI method remains a point for discussion, which cannot be fully covered in this work. 

Although the results presented here represent a useful baseline for NLSI benchmarks, the authors feel that it is important to stress that these should not be interpreted as `target' scores. % 
In particular, the authors would caution against peeking at the test scores during the algorithmic design stage. %
It is the intention of the authors that the results in this paper serve as a useful starting point for the reporting of baseline results. %
It is hoped that authors of new methodologies, benchmarks and applications in NLSI will be encouraged to contribute baselines alongside novel contributions. % 

\section{Conclusion}

In this paper, the authors make the case for the establishment of `baseline' results on benchmark datasets. %
It is argued that such baselines promote objective model comparison and defend against isolated model evaluation ensuring that advanced methodologies are compared against models from all classes. %
Results from ten baseline NLSI models from three model classes are presented on five publicly available benchmark datasets. %
The baseline results are compared to a selection of SOTA results from the literature. %
The baseline results offer strong performances in several benchmarks, however, they lag behind the SOTA by up to an order of magnitude in some cases. %

% \section*{Acknowledgements}

% MDC and TJR were supported by the EPSRC grant EP/W002140/1.

\bibliography{ifacconf}    

\begin{thebibliography}{27}
\providecommand{\natexlab}[1]{#1}
\providecommand{\url}[1]{\texttt{#1}}
\providecommand{\urlprefix}{URL }
\expandafter\ifx\csname urlstyle\endcsname\relax
  \providecommand{\doi}[1]{doi:\discretionary{}{}{}#1}\else
  \providecommand{\doi}{doi:\discretionary{}{}{}\begingroup \urlstyle{rm}\Url}\fi

\bibitem[{Akaike(1998)}]{akaike1998information}
Akaike, H. (1998).
\newblock Information theory and an extension of the maximum likelihood principle.
\newblock In \emph{Selected papers of {Hirotugu Akaike}}, 199--213. Springer.

\bibitem[{Beintema et~al.(2021)Beintema, Toth, and Schoukens}]{beintema2021nonlinear}
Beintema, G., Toth, R., and Schoukens, M. (2021).
\newblock Nonlinear state-space identification using deep encoder networks.
\newblock In \emph{Learning for dynamics and control}, 241--250.

\bibitem[{Beintema et~al.(2023)Beintema, Schoukens, and T{\'o}th}]{beintema2023continuoustime}
Beintema, G.I., Schoukens, M., and T{\'o}th, R. (2023).
\newblock Continuous-time identification of dynamic state-space models by deep subspace encoding.
\newblock In \emph{The Eleventh International Conference on Learning Representations}.

\bibitem[{Billings(2013)}]{Billings2013}
Billings, S.A. (2013).
\newblock \emph{Nonlinear System Identification: NARMAX Methods in the Time, Frequency, and Spatio-Temporal Domains}.
\newblock Wiley.

\bibitem[{Chung et~al.(2014)Chung, Gulcehre, Cho, and Bengio}]{chung2014empirical}
Chung, J., Gulcehre, C., Cho, K., and Bengio, Y. (2014).
\newblock Empirical evaluation of gated recurrent neural networks on sequence modeling.
\newblock In \emph{NIPS 2014 Workshop on Deep Learning, December 2014}.

\bibitem[{Enqvist and Ljung(2005)}]{Enqvist2005a}
Enqvist, M. and Ljung, L. (2005).
\newblock {Linear approximations of nonlinear FIR systems for separable input processes}.
\newblock \emph{Automatica}, 41(3), 459--473.

\bibitem[{Forgione and Piga(2021{\natexlab{a}})}]{forgione2021continuous}
Forgione, M. and Piga, D. (2021{\natexlab{a}}).
\newblock Continuous-time system identification with neural networks: Model structures and fitting criteria.
\newblock \emph{Eur. J. of Control}, 59, 69--81.

\bibitem[{Forgione and Piga(2021{\natexlab{b}})}]{forgione2021dynonet}
Forgione, M. and Piga, D. (2021{\natexlab{b}}).
\newblock dynonet: A neural network architecture for learning dynamical systems.
\newblock \emph{International Journal of Adaptive Control and Signal Processing}, 35(4), 612--626.

\bibitem[{Hochreiter and Schmidhuber(1997)}]{hochreiter1997long}
Hochreiter, S. and Schmidhuber, J. (1997).
\newblock Long short-term memory.
\newblock \emph{Neural computation}, 9(8), 1735--1780.

\bibitem[{Janot et~al.(2019)Janot, Gautier, and Brunot}]{Janot2019}
Janot, A., Gautier, M., and Brunot, M. (2019).
\newblock Data set and reference models of {EMPS}.
\newblock In \emph{2019 Workshop on Nonlinear System Identification Benchmarks}.

\bibitem[{Kingma and Ba(2014)}]{kingma2014adam}
Kingma, D.P. and Ba, J. (2014).
\newblock Adam: A method for stochastic optimization.
\newblock \emph{arXiv preprint arXiv:1412.6980}.

\bibitem[{Ljung(1999)}]{Ljung1999}
Ljung, L. (1999).
\newblock \emph{{System Identification: Theory for the User (second edition)}}.
\newblock Prentice Hall, Upper Saddle River, New Jersey.

\bibitem[{Murphy(2022)}]{pml1Book}
Murphy, K.P. (2022).
\newblock \emph{Probabilistic Machine Learning: An introduction}.
\newblock MIT Press.

\bibitem[{Paduart et~al.(2012)Paduart, Lauwers, Pintelon, and Schoukens}]{paduart2012identification}
Paduart, J., Lauwers, L., Pintelon, R., and Schoukens, J. (2012).
\newblock Identification of a wiener--hammerstein system using the polynomial nonlinear state space approach.
\newblock \emph{Control Engineering Practice}, 20(11), 1133--1139.

\bibitem[{Paduart et~al.(2010)Paduart, Lauwers, Swevers, Smolders, Schoukens, and Pintelon}]{paduart2010identification}
Paduart, J., Lauwers, L., Swevers, J., Smolders, K., Schoukens, J., and Pintelon, R. (2010).
\newblock Identification of nonlinear systems using polynomial nonlinear state space models.
\newblock \emph{Automatica}, 46(4), 647--656.

\bibitem[{Pascanu et~al.(2013)Pascanu, Mikolov, and Bengio}]{pascanu2013difficulty}
Pascanu, R., Mikolov, T., and Bengio, Y. (2013).
\newblock On the difficulty of training recurrent neural networks.
\newblock In \emph{International conference on machine learning}, 1310--1318.

\bibitem[{Pintelon and Schoukens(2012)}]{Pintelon2012}
Pintelon, R. and Schoukens, J. (2012).
\newblock \emph{{System Identification: A Frequency Domain Approach}}.
\newblock Wiley-IEEE Press, Hoboken, New Jersey, 2nd edition.

\bibitem[{Rasmussen and Williams(2006)}]{Rasmussen2006}
Rasmussen, C. and Williams, C. (2006).
\newblock Gaussian processes for machine learning.

\bibitem[{Relan et~al.(2017)Relan, Tiels, Marconato, and Schoukens}]{relan2017unstructured}
Relan, R., Tiels, K., Marconato, A., and Schoukens, J. (2017).
\newblock An unstructured flexible nonlinear model for the cascaded water-tanks benchmark.
\newblock \emph{IFAC-PapersOnLine}, 50(1), 452--457.

\bibitem[{Schoukens and Ljung(2009)}]{Schoukens2009}
Schoukens, J. and Ljung, L. (2009).
\newblock {Wiener-Hammerstein} benchmark.
\newblock In \emph{15th IFAC Symposium on System Identification}, 1--4.

\bibitem[{Schoukens and Ljung(2019)}]{Schoukens2019}
Schoukens, J. and Ljung, L. (2019).
\newblock Nonlinear system identification: {A} user-oriented road map.
\newblock \emph{IEEE Control Systems Magazine}, 39(6), 28--99.

\bibitem[{Schoukens et~al.(2016)Schoukens, Mattsson, Wigren, and No{\"e}l}]{SchoukensM2016}
Schoukens, M., Mattsson, P., Wigren, T., and No{\"e}l, J.P. (2016).
\newblock Cascaded tanks benchmark combining soft and hard nonlinearities.
\newblock Technical report, 4TU.ResearchData Dataset.

\bibitem[{Schoukens and No{\"e}l(2017)}]{SchoukensM2017}
Schoukens, M. and No{\"e}l, J.P. (2017).
\newblock Three benchmarks addressing open challenges in nonlinear system identification.
\newblock In \emph{20th World Congress of the International Federation of Automatic Control}, 448--453.

\bibitem[{Schuster and Paliwal(1997)}]{schuster1997bidirectional}
Schuster, M. and Paliwal, K.K. (1997).
\newblock Bidirectional recurrent neural networks.
\newblock \emph{IEEE transactions on Signal Processing}, 45(11), 2673--2681.

\bibitem[{Wigren and Schoukens(2013)}]{Wigren2013}
Wigren, T. and Schoukens, J. (2013).
\newblock Three free data sets for development and benchmarking in nonlinear system identification.
\newblock In \emph{2013 European Control Conference (ECC)}, 2933--2938.

\bibitem[{Wigren and Schoukens(2017)}]{Wigren2017}
Wigren, T. and Schoukens, M. (2017).
\newblock Coupled electric drives data set and reference models.
\newblock Technical report, Dep. of Information Technology, Uppsala University.

\bibitem[{Worden et~al.(2018)Worden, Barthorpe, Cross, Dervilis, Holmes, Manson, and Rogers}]{worden2018evolutionary}
Worden, K., Barthorpe, R., Cross, E., Dervilis, N., Holmes, G., Manson, G., and Rogers, T. (2018).
\newblock On evolutionary system identification with applications to nonlinear benchmarks.
\newblock \emph{Mechanical Systems and Signal Processing}, 112, 194--232.

\end{thebibliography}

\end{document}